# Eliminating the Electric Field Response in a Perovskite Heterojunction Solar Cell to Improve Operational Stability

Jiangjian Shi,[1] Yiming Li,[1] Yusheng Li,[1] Huijue Wu,[1] Yanhong Luo,[1,3,4] Dongmei Li,[1,3,4] Qingbo Meng[1,2,4*]

1. Key Laboratory for Renewable Energy, Chinese Academy of Sciences, Beijing Key Laboratory for New Energy Materials and Devices, Institute of Physics, Chinese Academy of Sciences, Beijing 100190, P. R. China
2. Center of Materials Science and Optoelectronics Engineering, University of Chinese Academy of Sciences, Beijing 100049, P. R. China
3. Department of Physics Science, University of Chinese Academy of Sciences, Beijing 100049, P. R. China
4. Songshan Lake Materials Laboratory, Dongguan, Guangdong 523808, P. R. China

* Corresponding author: qbmeng@iphy.ac.cn

**Abstract:** Intrinsic and extrinsic ion migration is a very large threat to the operational stability of perovskite solar cells and is difficult to completely eliminate due to the low activation energy of ion migration and the existence of internal electric field. We propose a heterojunction route to help suppress ion migration, thus improving the operational stability of the cell from the perspective of eliminating the electric field response in the perovskite absorber. A heavily doped p-type ($p^+$) thin layer semiconductor is introduced between the electron transporting layer (ETL) and perovskite absorber. The heterojunction charge depletion and electric field are limited to the ETL and $p^+$ layers, while the perovskite absorber and hole transporting layer remain neutral. The $p^+$ layer has a variety of candidate materials and is tolerant of defect density and carrier mobility, which makes this heterojunction route highly feasible and promising for use in practical applications.

**Introduction**

Owing to the outstanding semiconductor properties of organic-inorganic hybrid metal halides, perovskite solar cells have emerged as promising candidates for low-cost photovoltaic applications.[1-9] The power conversion efficiency (PCE) of the cell has reached 25.2%, comparable to that of a single crystal silicon solar cell.[10] The commercialization of perovskite solar cells is also in progress by developing large-scale film deposition, module production and tandem device techniques.[11-15] Nonetheless, the performance degradation of the cell under continuous operation has been a fundamental issue affecting the reliability of the product in practical outdoor applications.[16-18] This stability issue has been increasingly studied in both scientific and industrial fields.

After understanding the formation/decomposition mechanism of the perovskite material,[19-22] the humidity, high temperature and UV light exposure-related stability has been significantly improved through material and encapsulation engineering.[23-31] Recently, bias voltage- or light illumination-induced performance degradation has received increasing concern.[17-18,32-33] This issue is closely correlated to the intrinsic physics properties of perovskite absorbers, such as ion migration and defects, and is difficult to relieve by simple environmental isolation approaches. It has been found that both the bias voltage and light illumination can generate more defects in the cell by inducing ion (vacancy) migration/accumulation, lattice distortion or interfacial strain.[34-35] A series of works have demonstrated that ion migration is an unavoidable phenomenon in hybrid perovskites due to the relatively low activation energy of ion migration and the existence of an internal electric field.[34-39] As a self-doping multinary semiconductor, the defect and carrier properties of the perovskite are ultrasensitive to the ion (vacancy) distribution.[40] The long-distance ion migration and local accumulation can also induce lattice distortion and interfacial strain, thereby changing the atomic coordination of the interface.[41] Thus, ion (vacancy) migration/accumulation has a very large influence on the operational stability of the cell.

It was once reported that ion migration-induced cell performance degradation is reversible under day/night cycling.[42] However, irreversible performance loss was also observed under long-term day/night cycling operations, even under weak light intensities and moderate temperatures.[32] As a reverse process to degradation, performance recovery also depends on ion migration and restoration, which is a dynamic process and should be tightly correlated to the ambient temperature and recovery duration. In some sunny areas, such as the Gobi desert, the very large day-night temperature difference and low temperature at night cannot promote sufficient ion migration and restoration at night. For some areas in some seasons, the night duration is much shorter than the day duration. All these real-world conditions will more or less limit the degree of performance recovery, and thus, may cause permanent degradation. As such, in practical applications, we should not pin all our hope of sustaining the electricity production ability of the cells on the performance recovery phenomenon. Therefore, suppressing the intrinsic and extrinsic ion migration in the cell is highly necessary for improving cell stability. Ion migration mainly depends on the activation energy, available migration channels and internal electric field in the cell.[38] Numerous efforts, including alloying cations or anions, improving perovskite crystal quality, dimensional engineering and developing hole transporting layers (HTLs) without dopants, have been studied to enhance the activation energy of ion migration, reduce the available migration channels (for example, grain boundaries or point defects) or suppress the effect of extrinsic ions, respectively; thus far, positive effects have been achieved.[23,43-47]

Herein, we propose a heterojunction route to suppress ion migration and thus improve cell stability from the perspective of eliminating the internal electric field and field response in the perovskite absorber. A heavily p-type ($p^+$)-doped semiconductor layer is introduced between the n-type electron transporting layer (ETL) and the perovskite absorber to form an ETL-$p^+$-perovskite-HTL heterojunction. This structure eliminates the charge depletion, built-in electric field and electric field response in the perovskite absorber and thus can help stabilize the ion distribution of the perovskite. This heterojunction device structure can overcome the contradiction between reducing the internal electric field while sustaining a high PCE in a conventional ETL-perovskite-HTL device structure. Moreover, the $p^+$ semiconductor layer has an abundance of candidate materials and exhibits a high tolerance toward defect density and carrier mobility, which makes this heterojunction route highly feasible for use in practical devices. Thus, a promising approach to improve the operational stability of perovskite solar cells is provided.

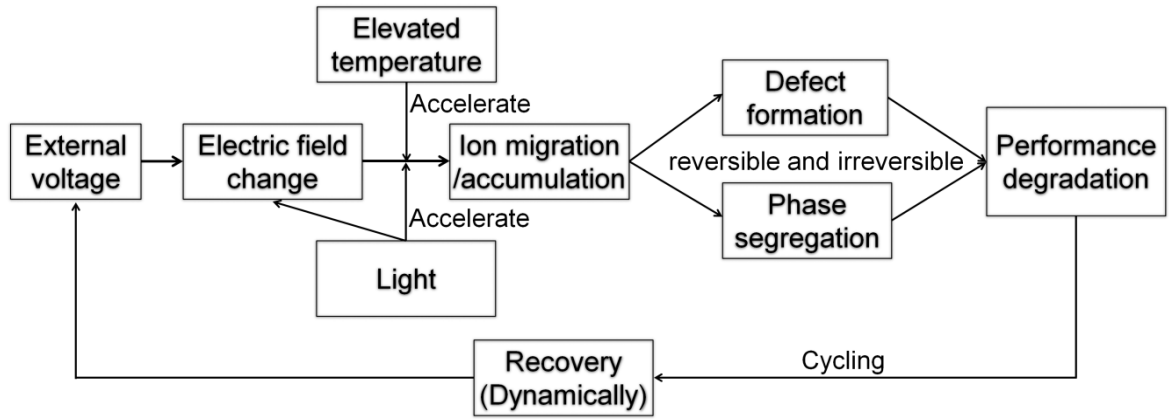

Scheme 1. The mechanism of the ion migration-induced operational stability issue of the perovskite solar cell.

**Results and discussions**

**Operational stability and ion migration in the cell**

Scheme 1 presents a logic diagram to exhibit the mechanism of the ion migration-related operational stability issue of the cell, which has been briefly discussed in the above section. Regarding the ion hopping and migration process, the electric field is the driving force, and the temperature and light illumination influence the dynamic parameters. The electric field is directly dependent on the external voltage and heterojunction properties. The external voltage is an important variable in a practical cell. In the dark (night), no external voltage is applied to the cell, while in the day (light), the external voltage at the cell can reach 1.0 eV in the maximum power point (MPP) tracking mode. When the sunlight intensity changes during the day, the external voltage will fluctuate accordingly. The change in electric field during the day/night can reach $10^4$ V cm$^{-1}$, which is large enough to drive ion migration and redistribution.[48] During the whole life of an outdoor cell, this will occur thousands of times. Ion migration is in fact a common phenomenon in photovoltaic devices, such as Cu(In, Ga)Se$_2$ and CdTe solar cells,[49] and in energy storage devices, such as lithium- and sodium-ion batteries. In these devices, ion migration has little influence on the stability of the material framework. In contrast, mobile ions such as methylamine (MA), iodine and bromine ions in the hybrid perovskite have a large influence on the lattice structure stability, defect composition and local phase.

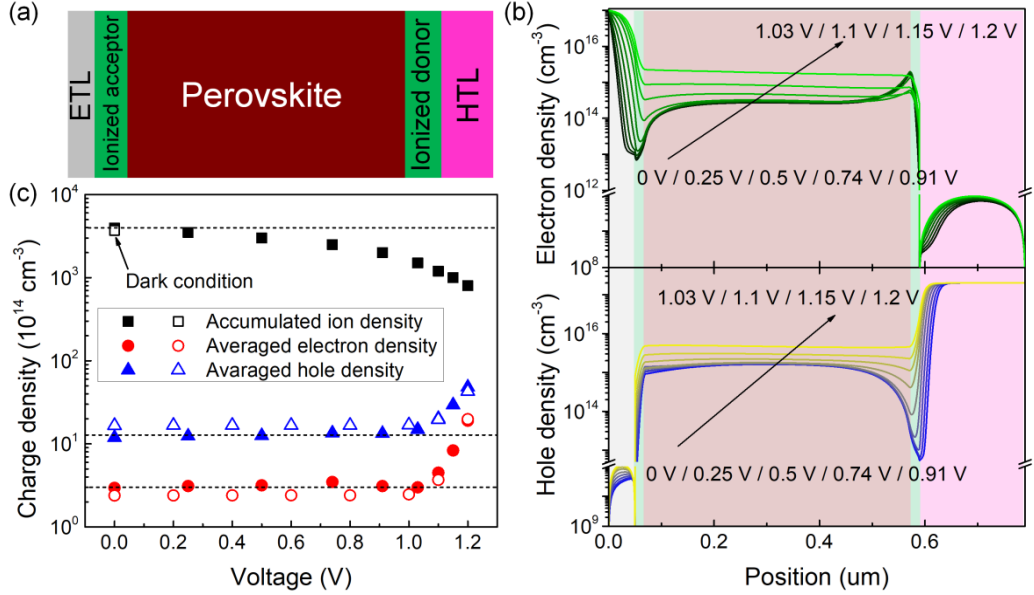

Figure 1. Ion accumulation and excess carriers in the cell. (a) Device model for the perovskite solar cell that has ion migration and accumulation. (b) Steady-state distribution of excess free electrons and holes in the cell under different steady-state voltages. In the steady-state condition, ions (vacancies) have accumulated at the interface to screen bulk electric field. (c) Steady-state voltage dependent densities of averaged excess carriers (open scatters for our proposed heterojunction structure discussed later) of the perovskite and accumulated ions (vacancies) at the interface.

To gain clear insight into the ion migration phenomenon in the cell, we first performed a steady-state device simulation of the ion accumulation and carrier distribution in the cell at various bias voltages. The well-known microelectronic and photonic structure analysis program AMPS based on Poisson's equation and drift-diffusion model was used for all the device simulations in this work. This program has been widely applied in heterojunction device simulations, and its reliability has been shown in previous studies.[50-52] The device model used here is depicted in Figure 1(a), and more detailed parameters are summarized in Table S1. The perovskite absorber (500 nm) is sandwiched by the ETL (50 nm) and the HTL (200 nm). To reflect ion accumulation under an internal electric field, thin layers (20 nm) comprised of high-density ionized acceptors (negative charge, such as $V_{MA}$, $I_i$ and $I_{MA}$) and donors (positive charge, such as $MA_i$, $V_I$ and $MA_I$) are introduced at the ETL/perovskite and perovskite/HTL interfaces, respectively.[40,53-55] Immediately after the cell is fabricated, ion migration and accumulation should have already occurred due to the built-in electric field of the heterojunction.[55] Under steady-state conditions, the electric field in the bulk perovskite is screened by ion accumulation at the interface, and the cell provides a steady-state PCE of >21% (Figure S1). Due to the ion screen effect, the carrier distribution in the perovskite absorber almost does not change at voltages <1.1 V, as presented in Figure 1(b).

Compared to the carrier density, ion migration/accumulation is a more significant variable in the cell, as exhibited by the bias voltage-dependent charge density in Figure 1(c). The density of accumulated ions at the interface ($N_I$) can reach ~$3.6 \times 10^{17}$ cm$^{-3}$ in an as-fabricated cell. When light illumination is applied, ion migration will continue to a higher $N_I$ of ~$4.0 \times 10^{17}$ cm$^{-3}$. Under a high voltage of 1.0 V (close to the MPP), the $N_I$ is ~$1.5 \times 10^{17}$ cm$^{-3}$. This result is much lower than that in the short-circuit condition, and thus, would release the accumulated ions at the interface with a value of ~$2 \times 10^{17}$ cm$^{-3}$ (i.e., $4 \times 10^{11}$ cm$^{-2}$). This large ion redistribution will affect the stability of both the perovskite lattice framework and the interfacial structure and thus can cause significant cell performance degradation. At a higher voltage, such as in the open-circuit condition, an increase in ion redistribution (~$4.0 \times 10^{17}$ cm$^{-3}$) together with an increased carrier density can induce more severe cell performance degradation, as observed in the experiment.[56] In practical applications, the cell usually works close to the MPP or is stored in the dark. Under these conditions, ion migration is a more prominent factor that needs to be considered for improving cell stability.

**Electric field properties of the cell**

Ion migration is driven by the electric field or the electric field change in the perovskite absorber. In the cell, a heterojunction electric field is formed because of the different Fermi energy levels between the perovskite absorber, ETL and HTL, and the change in electric field is caused by the external voltage. Figure 2(a) shows a schematic diagram of this phenomenon in a standard ETL-perovskite-HTL three-layer-structured device. Due to charge depletion, the built-in electric field in the perovskite absorber is as high as $10^4$ V cm$^{-1}$; additionally, when the cell is switched to an operating condition (for example, 0.9 V under light illumination), the change in electric field can reach $10^4$ V cm$^{-1}$. As a response to this change in electric field, ion migration and redistribution will occur. Apparently, eliminating the electric field response

(including both the built-in electric field of the heterojunction and the change in electric field) in the perovskite is a direct route to suppress ion migration.

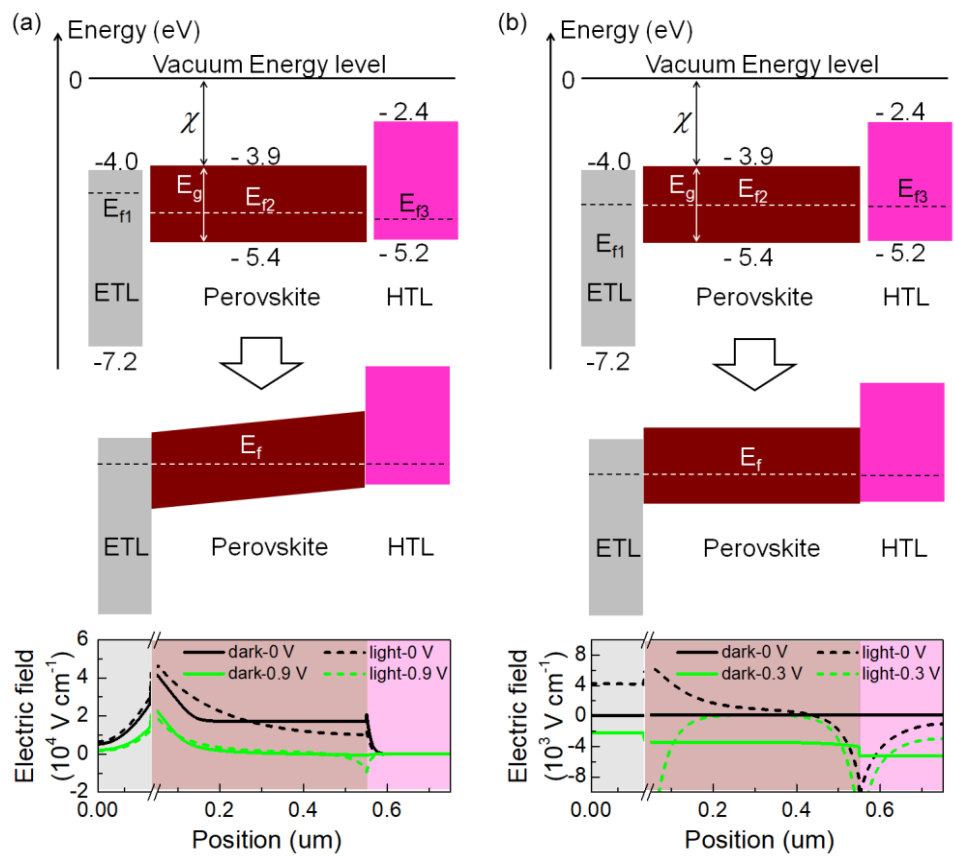

Figure 2. Energy band alignment and internal electric field distribution of a standard three-layer-structured perovskite solar cell. (a) Standard device: Fermi energy level of the isolated ETL, perovskite and HTL is different from each other. (b) Flat-band device: The isolated ETL, perovskite and HTL have the same Fermi energy level.

As an initial attempt, we adjusted the Fermi energy levels of the ETL, perovskite and HTL to be similar to each other. In the experiment, this adjustment is able to be realized by adjusting the doping properties of the perovskite and charge transporting layers (CTLs). This adjustment can help to completely eliminate the electric field of the heterojunction in the as-fabricated cell, as shown in Figure 2(b). However, when light illumination and a positive bias voltage (for example, 0.3 eV, as shown here) are applied to the cell, a negative electric field can be easily generated across the whole cell, which blocks charge extraction and transport, thus resulting in a low PCE of <5%. The ETL and HTL also exhibit a very large electric field response, which will induce extrinsic ion migration in the cell.[57] Thus, decreasing the difference in the Fermi energy levels hardly helps to improve the operational stability of the cell and is also unacceptable for practical applications because of the low PCE.

In a single heterojunction, the electric field response is determined by the built-in electric potential distribution in the cell. The electric potential in the light absorber side can be written as the following (take the ETL-perovskite (p-type) as an example):

$$V_D^{pero} = \frac{\varepsilon_{ETL} N_D^{ETL}}{q\left(\varepsilon_{pero} N_A^{pero} + \varepsilon_{ETL} N_D^{ETL}\right)} \left(CBO + E_g^{pero} + KT \ln \frac{N_D^{ETL} N_A^{pero}}{N_C^{ETL} N_V^{pero}} - qV\right), \quad (1)$$

where $V_D^{pero}$ is the electric potential at the perovskite layer, $\varepsilon_{ETL}$ and $\varepsilon_{pero}$ are the dielectric constants of the ETL and perovskite, respectively, $N_D^{ETL}$ and $N_A^{pero}$ are the donor and accepter concentrations of the ETL and perovskite, respectively, CBO is the conduction band offset defined as $E_C^{ETL}$-$E_C^{pero}$, $E_g^{peros}$ is the bandgap of the perovskite absorber, $N_C^{ETL}$ and $N_V^{pero}$ are the effective density of states of the conduction band and valence band of the ETL and perovskite, respectively, $q$ is the elementary charge, $K$ is the Boltzmann constant, $T$ is the thermodynamic temperature and $V$ is the external voltage.

The electric potential change ($\Delta V_D^{pero}$) caused by the external voltage can then be derived as

$$\Delta V_D^{pero} = \frac{\varepsilon_{ETL} N_D^{ETL} V}{\left(\varepsilon_{pero} N_A^{pero} + \varepsilon_{ETL} N_D^{ETL}\right)}. \quad (2)$$

In the cell, the averaged electric field and its change are proportional to $V_D^{pero}$ and $\Delta V_D^{pero}$, respectively. Equations (1) and (2) provide us with clear evidence that the electric field response of the perovskite absorber is mainly influenced by the CBO and the doping density of both the ETL and perovskite. When the HTL is also involved, its doping density and the valence band offset (VBO: $E_V^{HTL}$-$E_V^{pero}$) of the HTL/perovskite interface will also influence the electric field response.

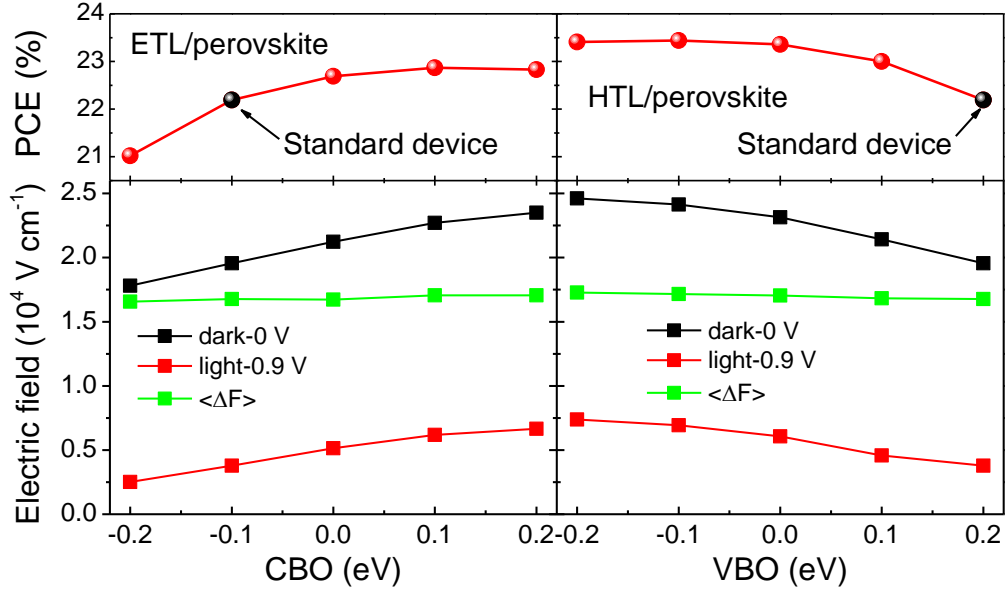

Figure 3. Influence of conduction or valence band offset (CBO: ETL/perovskite, VBO: HTL/perovskite) of the standard cell (without considering ion migration and screen effect) on the PCE, the averaged internal electric field strength (absolute strength, <$F$>) and the electric field strength change (<$\Delta F$>) in the perovskite layer.

According to equations (1) and (2), we explored the possibility of adjusting the CBO or VBO in suppressing the electric field response. The electron affinity of the ETL and HTL was changed for this purpose. In the experiment, this adjustment could be realized by various approaches, such as introducing ion doping or interfacial dipoles, changing the CTL film deposition conditions, using a plasma treatment or modifying the chemical groups of organic CTLs. The PCE, electric field and change in electric field of the cells with different CBOs or VBOs were calculated and are presented in Figure 3. For clarity, the electric field in the perovskite absorber has been averaged as

$$\langle F \rangle = \frac{1}{L} \int_{pero} |F(x)| dx, \qquad (3)$$

where <$F$> is the averaged electric field strength, $|F(x)|$ is the field strength at a certain position $x$, and $L$ is the thickness of the perovskite absorber. The electric field change ($\Delta F$) in the cell when it is switched from dark (0 V) to light (0.9 V) has also been averaged as

$$\langle \Delta F \rangle = \frac{1}{L} \int_{pero} |F(dark-0V) - F(light-0.9V)| dx. \qquad (4)$$

The standard cell exhibits a PCE of 22.2%, and <ΔF> reaches up to $1.7 \times 10^4$ V cm$^{-1}$. For the ETL/perovskite, when the CBO is increased to a positive value of 0.1 eV, a high PCE of 22.9% can be obtained. However, this PCE improvement is realized by strengthening the electric field. In addition, no obvious reduction in <ΔF> can be observed. For the HTL/perovskite, a high PCE of 23.4% can be obtained at a negative VBO of -0.1 eV. Similar to the ETL, adjusting the band-edge position of the HTL cannot reduce <ΔF> and, thus, is hardly able to help improve cell stability.

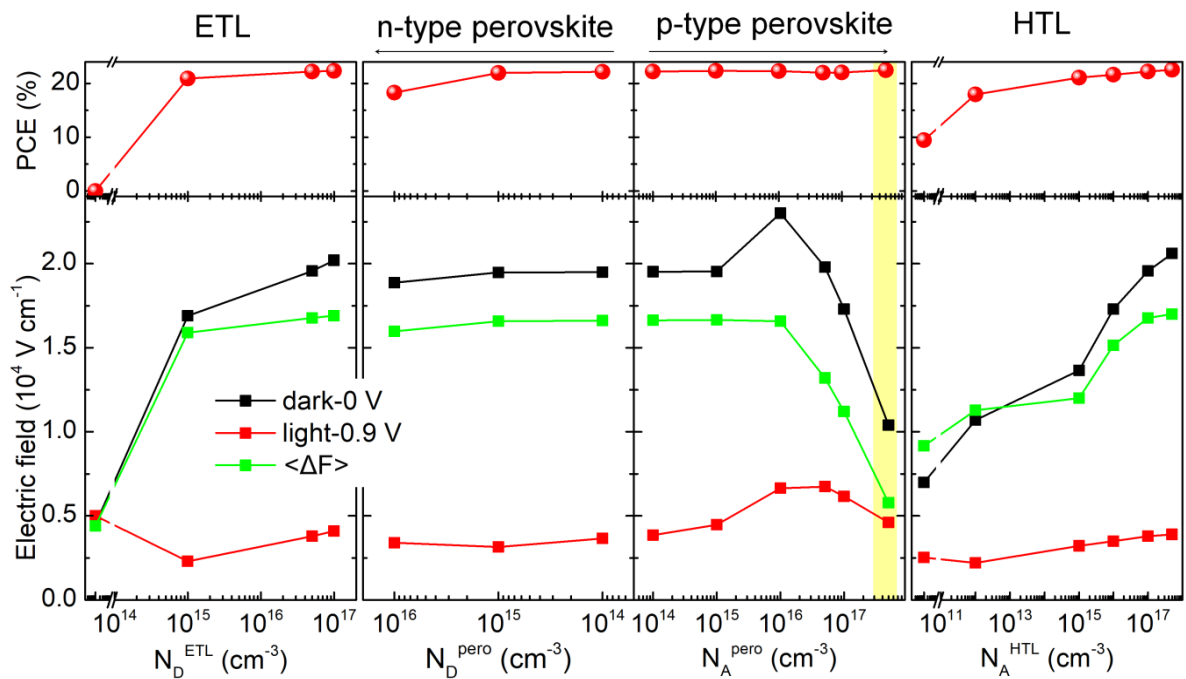

Figure 4. Influence of donor ($N_D$) or acceptor ($N_A$) doping density of the ETL, perovskite and HTL in the standard cell on the PCE, <F> and <ΔF> (without considering ion migration and screen effect). Only with high p-type doping the perovskite layer, can <ΔF> be obviously suppressed while sustaining a high PCE (yellow shadow).

The influence of the doping properties of each layer in the cell is further investigated over a wide concentration, as presented in Figure 4. For the ETL, the $N_D^{ETL}$ is adjusted from $10^{17}$ cm$^{-3}$ to 1 cm$^{-3}$. Decreasing the $N_D^{ETL}$ can help reduce the electric field response of the perovskite absorber. The <ΔF> value is suppressed to <$0.5 \times 10^4$ V cm$^{-1}$ when the $N_D^{ETL}$ reaches 1 cm$^{-3}$. However, almost no power conversion can be obtained at an $N_D^{ETL}$ that is this low. For the HTL, a similar trend is observed. When the acceptor concentration of the HTL ($N_A^{HTL}$) is decreased to $10^5$ cm$^{-3}$, $<\Delta F>$ can be reduced to $<1.0 \times 10^4$ V cm$^{-1}$ but results in a low PCE of ~10%. This decrease in PCE is due to the weakening of the heterojunction potential and the charge transporting ability of the CTLs. The electric field in the ETL or HTL can be easily converted to a negative value by external voltages, which will greatly reduce their charge extraction and collection ability. As such, to sustain desirable charge output at high voltages, the cell needs to have a high heterojunction potential.

For the perovskite absorber, both n-type and p-type doping properties were considered. Figure 4 shows that the cell can sustain a high PCE over a wide doping range, from moderate n-type ($N_D^{pero}$: $10^{15}$ cm$^{-3}$) to high p-type doping ($N_A^{pero}$: $5 \times 10^{17}$ cm$^{-3}$). This observation agrees well with the experimental results that both n- and p-type perovskites can provide efficient solar energy conversion.[58-59] The $<\Delta F>$ value is reduced to $0.6 \times 10^4$ V cm$^{-1}$ when the $N_A^{pero}$ is increased to $5 \times 10^{17}$ cm$^{-3}$. This result agrees with equations (1) and (2) in that a high $N_A^{pero}$ can help reduce the electric potential depletion and electric field response in the light absorber. At this $N_A^{pero}$, the perovskite exhibits a wide neutral region of more than 450 nm, in which ion migration can be suppressed. In this neutral region, the carriers transport through a diffusion mechanism, and the long carrier diffusion length of the perovskite ensures efficient charge transport/extraction and thus a high PCE of 22.4%. This result means that an electric field in the bulk perovskite is not necessary for obtaining a high PCE. Nonetheless, the perovskite still exhibits a high electric field of $>10^5$ V cm$^{-1}$ at both the ETL/perovskite and HTL/perovskite interfaces, and the change in electric field at the ETL/perovskite interface can reach $10^5$ V cm$^{-1}$ (Figure S2). This result may still induce ion migration and accumulation. Moreover, the high p-type doping of the perovskite is usually realized by a high concentration of self-doping point defects, such as cation vacancies and anion/cation substitutions. This high concentration of point defects may provide more ion migration channels. In addition, the existence of deep defects or lattice distortion can also be induced by the high concentration of self-doping point defects, thereby influencing the PCE. In summary, although a high $N_A^{pero}$ can help theoretically eliminate the bulk electric field and ion migration, it may not be an acceptable method in practice.

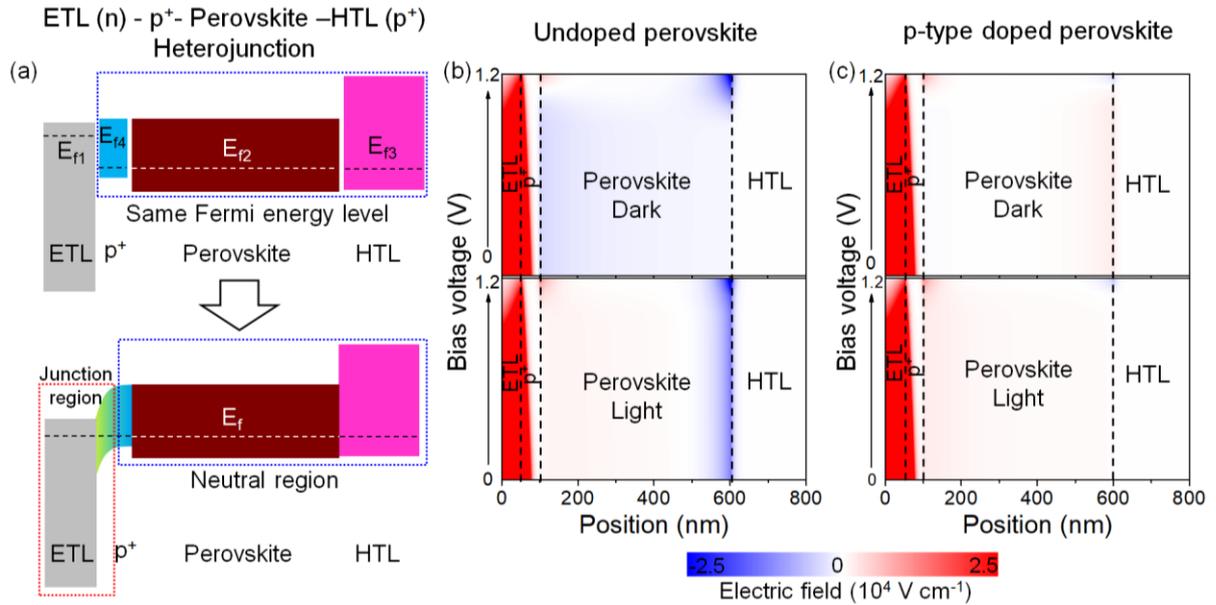

Figure 5. A new heterojunction structure to eliminate the electric field response in the perovskite layer. (a) Schematic diagram of the device structure and energy band bending of the newly proposed heterojunction cell. Compared to the standard cell, a high p-type-doped ($p^+$) semiconductor layer is introduced between the ETL and the perovskite absorber, forming a four-layer-structured device, to suppress the charge depletion and internal electric field in the perovskite absorber. (b-c) Bias voltage dependent electric field distribution of the cell in dark and under light illumination. (b. the perovskite is not doped; c. the perovskite is doped with $10^{15}$ cm$^{-3}$ acceptors)

**Four-layer-structured heterojunction for eliminating the internal electric field**

Although adjusting the heterojunction properties of a standard structured cell can hardly improve the operational stability of the cell while sustaining a high PCE, the above studies still provide several valuable results regarding the electric field and cell performance. First, a large heterojunction depletion electric potential is necessary for obtaining a high PCE. Second, an electric field in the bulk light absorber is not always necessary for obtaining a high PCE as long as the light absorber has a long carrier diffusion length. Third, the electric field response in the bulk light absorber can be eliminated by limiting the charge depletion of the heterojunction at the interfacial region.

These results guide us to design a new heterojunction device structure, as shown in Figure 5(a). A thin (50 nm as an example) and high p-type-doped semiconductor layer (denoted as $p^+$ layer) is introduced between the ETL and perovskite. The Fermi energy level of this $p^+$ layer is similar to that of the perovskite absorber but has a large difference from that of the ETL. A depleted heterojunction is thus formed between the ETL and the $p^+$ layer, and the perovskite absorber can remain neutral, as depicted in Figure 5(a). The electric field response caused by the heterojunction formation and external bias voltage will be limited to the ETL and $p^+$ layer. As such, the ion distribution in the perovskite absorber can be stabilized. The HTL/perovskite junction can also be modified by this route. Here, we mainly focus on the ETL/perovskite interface because we have found that the electric field response mainly exists at this interface. The electric field profile of this four-layer-structured heterojunction cell and its response to an external bias voltage and light illumination are quantified by device simulation, and an example (Table S2) is shown in Figure 5(b-c). Both the intrinsic and p-type ($N_A^{pero} = 10^{15}$ cm$^{-3}$) perovskites are considered here.

It can be clearly seen that an electric field mainly exists in the ETL and $p^+$ layer with a field strength reaching $10^5$ V cm$^{-1}$. For the intrinsic perovskite, a weak negative electric field (~ -$10^3$ V cm$^{-1}$) exists in the cell in the dark, and the field exhibits a negligible response to the external bias voltage (<1.0 V). When light illumination is applied, the bulk electric field is reversed to ~+$10^2$ V cm$^{-1}$. In addition, a large negative electric field of $10^4$ V cm$^{-1}$ appears at the HTL/perovskite interface and even extends into the HTL, which may induce intrinsic or extrinsic ion migration. This behavior arises because the photoinduced carriers (~$10^{14}$ cm$^{-3}$) have changed the charge distribution in the intrinsic perovskite absorber. Comparatively, when the perovskite is doped, the light illumination has little influence on the electric field. Regardless of whether the cell is in the dark or under illumination, only a weak positive electric field of $10^2$ V cm$^{-1}$ exists in the perovskite absorber and remains unchanged across the entire studied voltage range (0-1.2 V). More importantly, the electric field at the HTL/perovskite interface and in the HTL has been completely eliminated, thereby helping suppress the extrinsic ion drift from the HTL into the perovskite. Thus, our proposed heterojunction device should be able to help significantly relieve the ion migration-related stability issue of the cell.

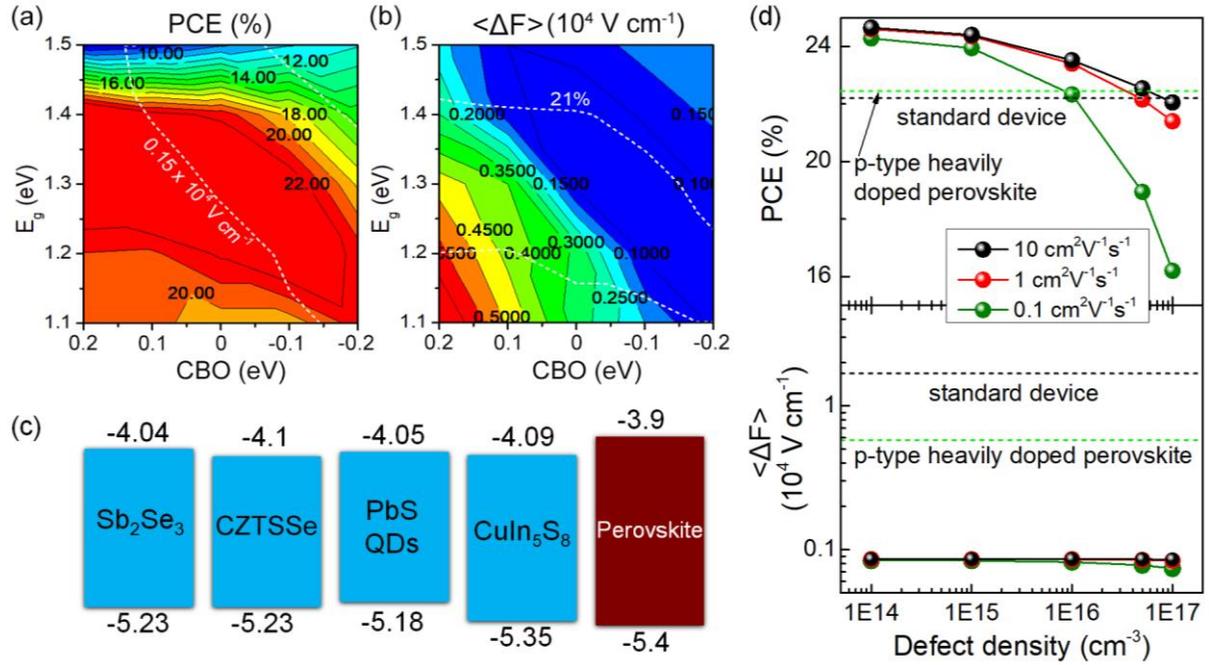

Figure 6. Material search for the p$^+$ layer. (a-b) Two dimensional pseudocolor plots of PCE and <$\Delta F$> for the new heterojunction device as functions of CBO and $E_g$ of the p$^+$ semiconductor. The white dashed lines in (a) and (b) mark out the high PCE (>21%) or low <$\Delta F$> (<0.15×10$^4$ V cm$^{-1}$) regions, respectively. (c) Possible semiconductor materials for the p$^+$ layer. (d) Influence of defect density and carrier mobility of the p$^+$ semiconductor on the PCE and <$\Delta F$>.

To evaluate the possibility of constructing this heterojunction device in practice, we further search for a material to use as the critical p$^+$ layer. We systematically regulate the band-edge position (reflected by the CBO) and bandgap ($E_g$) of the p$^+$ semiconductor and calculate their influence on the PCE and <$\Delta F$> of the cell, as presented in Figure 6(a-b). On the one hand, the cell can obtain a higher PCE than the standard cell in a wide CBO-$E_g$ region (marked by the contour plot labeled 22 (red region) in Figure 6(a)). This result arises because the p$^+$ layer has significantly suppressed electric response of the perovskite absorber and thus reduced the interfacial charge recombination at high bias voltages (Figures S3 and S4). On the other hand, the cell exhibits significantly reduced <$\Delta F$> in the entire studied CBO-$E_g$ region (CBO: -0.2~0.2 eV and $E_g$: 1.1~1.5 eV), as shown in Figure 6(b). For example, the contour plot labeled 0.15 (blue region) means that <$\Delta F$> is 0.15×10$^4$ V cm$^{-1}$, which is ten times lower than that of the standard cell.

More importantly, we can find an overlapping region between the high PCE and low $<\Delta F>$. For example, the white dashed lines in Figure 6(a) and (b) mark out a region where the cell provides a PCE of >21%, while its $<\Delta F>$ is smaller than $0.15 \times 10^4$ V cm$^{-1}$. According to previous works,[48,60-61] we believe $<\Delta F>$ at this level will not cause obvious cell performance degradation. Although the PCE is slightly lower than that of the standard cell, this small sacrifice in PCE is worthwhile if the operational stability of the cell can be significantly improved. According to previously experimental reports, we can find several common materials suitable for the p$^+$ layer, such as Sb$_2$Se$_3$, Cu$_2$ZnSn(S, Se)$_4$ (CZTSSe), PbS quantum dots (QDs) and CuIn$_5$S$_8$, as depicted in Figure 6(c).[62-65] Further from a material database based on first-principles calculations,[66] more than one hundred semiconductors with $E_g$ in the range from 1.15 to 1.4 eV can be considered as potential candidate materials, and some examples are given in Table S3. In practice, both the band-edge position and $E_g$ of the perovskite are tunable, which makes more materials usable for the p$^+$ layer. Thus, our proposed heterojunction device structure has a large tolerance to the energy band of the p$^+$ layer.

The influence of defect density (from $10^{14}$ cm$^{-3}$ to $10^{17}$ cm$^{-3}$) and carrier mobility (from 0.1 to 10 cm$^2$V$^{-1}$s$^{-1}$) of the p$^+$ layer on the PCE and $<\Delta F>$ has also been considered, as shown in Figure 6(d). Here, device parameters similar to those in Figure 5(c) are used. It is found that $<\Delta F>$ (~$0.08 \times 10^4$ V cm$^{-1}$) is almost independent of the defect density and carrier mobility. The cell can always sustain a high PCE of >21% as long as the p$^+$ layer has a moderate carrier mobility (for example ≥ 1 cm$^2$V$^{-1}$s$^{-1}$). This demand can be easily satisfied by Sb$_2$Se$_3$, CZTSSe and CuIn$_5$S$_8$.[60,67-68] Although a PbS QD film usually has a low carrier mobility, its defects can be effectively suppressed by ligand passivation, and the carrier lifetime can reach tens of microseconds.[64] As a result, the carrier transport and recombination properties of the PbS QD film will not limit its use as a p$^+$ layer. Therefore, these high tolerances of the p$^+$ layer toward the energy band, defect density and carrier mobility make our proposed heterojunction structure feasible for use in practical devices.

**Conclusions**

In this work, we proposed a four-layer-structured perovskite heterojunction (ETL/$p^+$/perovskite/HTL) route for eliminating the electric field response in the perovskite absorber, which is expected to suppress the intrinsic or extrinsic ion migration, thus enhancing the operational stability of the perovskite solar cell. Compared to a conventional perovskite cell, a thin $p^+$ layer has been introduced in this new heterojunction device structure. The junction charge depletion and built-in electric field mainly exist in the ETL and $p^+$ layer, while the perovskite absorber and HTL are maintained as neutral. The external bias voltage and light illumination that the cell has to work under will no longer induce ion migration in the perovskite absorber. Although the photogenerated carriers in the neutral perovskite absorber transport mainly through the diffusion mechanism, owing to the long carrier diffusion length and suppressed interfacial charge recombination, the cell can still obtain an impressively high PCE. Therefore, this new heterojunction device structure has overcome the contradiction between reducing the internal electric field and sustaining a high PCE in a conventional perovskite cell. Through a material search, we further identify that the critical p+ layer has a variety of candidate materials and is highly tolerant of defect density and carrier mobility. Therefore, this heterojunction device is feasible for use in practical applications, thus providing a promising route for improving the operational stability of perovskite solar cells.

## Conflicts of interest

There are no conflicts to declare.

## Acknowledgements

This work was supported by the National Key R&D Program of China (no. 2018YFB1500101), the Natural Science Foundation of China (nos. 11874402, 51627803, 51421002, 91733301, 51761145042 and 51872321), and the International Partnership Program of Chinese Academy of Sciences (no. 112111KYSB20170089).